# Large enhancement in thermal conductivity of solvent cast expanded-graphite/polyetherimide composites

Fatema Tarannum[1], Swapneel Danayat[1], Avinash Nayal[1], Rajmohan Muthaiah[1], Jivtesh Garg*[1]

[1]School of Aerospace and Mechanical Engineering, University of Oklahoma, Norman, 73019, USA

**Abstract:**

We demonstrate in this work, that expanded graphite (EG) can lead to a very large enhancement in thermal conductivity of polyetherimide-graphene and epoxy-graphene nanocomposites prepared via solvent casting technique. A $k$ value of 6.56 Wm$^{-1}$K$^{-1}$ is achieved for 10 weight % composition sample, representing an enhancement of ~2770% over pristine polyetherimide ($k$ ~ 0.23 Wm$^{-1}$K$^{-1}$). This extraordinary enhancement in thermal conductivity is shown to be due to a network of continuous graphene sheets over long length scales, resulting in low thermal contact resistance at bends/turns due to the graphene sheets being covalently bonded at such junctions. Solvent casting offers the advantage of preserving the porous structure of expanded graphite in the composite, resulting in the above highly thermally conductive interpenetrating network of graphene and polymer. Solvent casting also does not break down the expanded graphite particles, due to minimal forces involved, allowing for efficient heat transfer over long length scales, further enhancing overall composite thermal conductivity. Comparisons with a recently introduced effective medium model shows a very high value of predicted particle-particle interfacial conductance, providing evidence for efficient interfacial thermal transport in expanded graphite composites. Field Emission Environmental Scanning Electron Microscopy (FE-ESEM) is used to provide detailed understanding of interpenetrating graphene-polymer structure in the expanded graphite composite. These results open up novel avenues for achieving high thermal conductivity polymer composites.

# 1. Introduction

Increased thermal dissipation in modern electronic devices has led to a demand for thermally conductive materials with superior thermal conductivity ($k$)[1,2]. Light weight, high corrosion resistance, and excellent processability of polymeric materials make them attractive for thermal management applications[3,4]. However, poor thermal conductivity (<0.5 Wm$^{-1}$K$^{-1}$)[5,6] of polymers limits their application in efficient heat removal process. Addition of high thermal conductivity fillers such as graphene or carbon nanotubes, has been shown to significantly enhance thermal conductivity of polymer-graphene nanocomposites[7-13]. Different effects have been investigated to enhance the effectiveness of graphene in enhancing $k$ such as alignment and percolation effects. Alignment of carbon nanomaterials[14-20] takes advantage of their thermal conductivity along one direction (in-plane for graphene nanoplatelets and along axis for carbon nanotubes). However, the enhancement achieved in $k$ through alignment effects is anisotropic, potentially limiting the application of such composites. One of the most promising approaches has been the percolative effect where graphene-graphene contact is used to bypass the polymer, leading to significantly higher thermal conductivity enhancement[13,21,22]. More recently expanded graphite has been shown to yield very high composite $k$ values[23-28]. In this paper, we explore the use of expanded graphite (EG) for enhancing thermal conductivity of polyetherimide (PEID) due to the unique applications of this polymer in electrical systems[29-31], which can benefit from efficient heat dissipation. In particular we show that expanded-graphite/polymer composites prepared through solvent casting can lead to more efficient heat transfer due to an almost continuous network of graphitic sheets over long length scales, which overcomes the problem of low interfacial thermal conductance at graphene-graphene contact in typical percolation environments. We have measured a composite thermal conductivity of 6.6 Wm$^{-1}$K$^{-1}$ at just 10 wt% filler concentration. This value represents the highest $k$ value measured in a polymer composite at this low weight fraction of filler. Presented results provide new avenues for achieving efficient thermal management in a wide range of applications.

Intercalation of acid molecules and oxidizing agents into graphite (Fig. 1c) followed by rapid heating (900 °C) leads to a conversion of intercalation agents into a gaseous state (Fig. 1a) resulting in an expansion of graphite structure. This results in an increase in interlayer spacing yielding expanded graphite with a worm-like structure (Fig. 1d). EG achieved through such expansion has a highly porous (Fig. 1e and f), lightweight structure with a very low density (0.002–0.02 g/cm$^3$) and exhibits high mechanical strength (10 MPa), thermal conductivity (25–470 Wm$^{-1}$K$^{-1}$), electrical conductivity (10$^6$–10$^8$ S/cm)[32]. As a result EG has emerged as a promising material with applications such as flame retardancy[33], phase-change material[34, 35], electrodes,[36, 37] electrochemical sensors,[38] fuel cells,[39, 40], batteries[41, 42] and supercapacitors[43, 44].

Significant research has been performed into the use of EG in enhancing thermal conductivity of polymer-graphene nanocomposites. Wu[45] et al. measured thermal conductivity of individual expanded graphite particles using a T-type method and reported a value ~ 335 Wm$^{-1}$K$^{-1}$ for EG particles. Tao et al.[46] prepared EG/PDMS composite using a hot-press technique and reported a high thermal conductivity value of 4.7 Wm$^{-1}$K$^{-1}$ at 10 wt% EG composition. Zhao et al.[47] measured a high thermal conductivity of 3.5 Wm$^{-1}$K$^{-1}$ in EG/paraffin wax composites at 25 wt% composition. Song et al.[48] measured a thermal conductivity of 1.66 Wm$^{-1}$K$^{-1}$ at 20 wt% composition in EG/ MgCl$_2$·6H$_2$O composite. Wei et al.[49] created a network of expanded graphite particles using stearic acid and polyethylene wax and measured a high thermal conductivity of 19.6 Wm$^{-1}$K$^{-1}$ using ~25 vol% expanded graphite. Che et al.[50] employed synergistic effects between expanded graphite and carbon nanotubes to achieve a high thermal conductivity of ~3.0 Wm$^{-1}$K$^{-1}$ at ~20 wt% composition in high-density polyethylene (HDPE) composites.

In this work, we show that expanded graphite can lead to remarkable increase in thermal conductivity of polyetherimide composites prepared via the approach of solvent casting. Expanded graphite has a highly porous structure with interconnected graphitic walls as seen in the SEM image of expanded graphite in air in Fig. 1f. The graphitic sheets form a continuous network over long length scales, allowing efficient conduction of heat. The approach of solvent casting offers the advantage of preserving this porous structure of expanded graphite as seen in the Field Emission Environmental Scanning Electron Microscopy (FE-ESEM) images of expanded graphite, while it is embedded in the polymer composite, in Figs. 1h, i and k. Comparison of Figs. 1f (showing EG in air) and Figs. 1h, i and k (showing EG in the composite), clearly show that the porous structure of EG is retained within the composite. This is due to the use of solvent casting, which does not exert large forces on expanded graphite, preserving its porous structure, unlike micro-compounding where large shear forces distort this structure. Such a porous structure gets infused with the dissolved polymer during the casting process resulting in a highly conductive interpenetrating network of graphene and polymer.

The resulting interpenetrating polymer-graphene structure from the solvent casting method allows for very efficient heat transfer. This can be seen by observing that graphitic nanosheets in this network are continuous for very long length scales (the lateral dimensions of these expanded

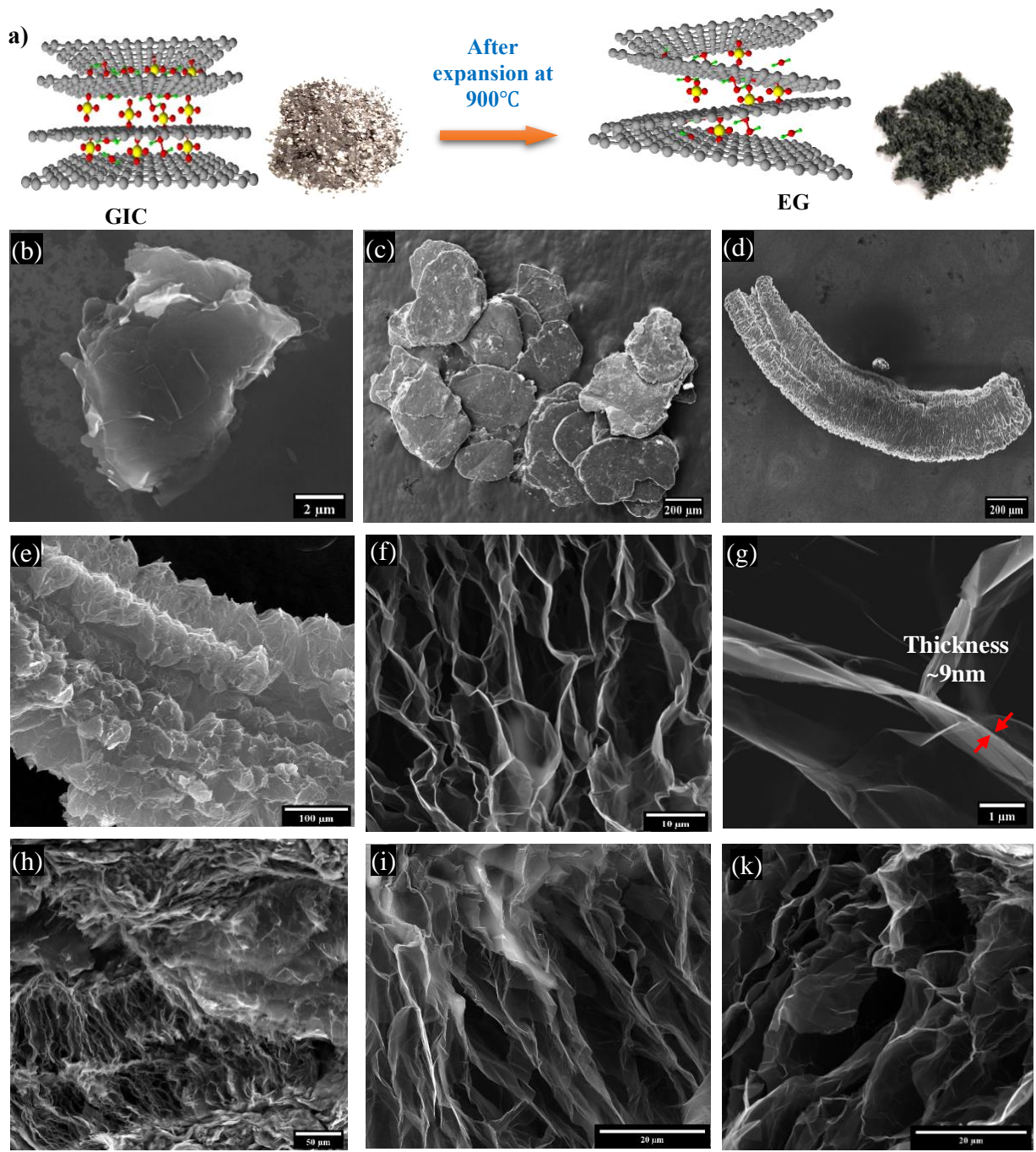

Figure 1 a) Expansion of intercalated graphene compound at 900 °C. SEM image of b) 60nm GnP, c) GIC d) EG after thermal expansion. High resolution images of EG at e) 100X, f) 1500X, g) 12000X magnification. FE-ESEM images of EG/PEID composite at h) 350x, (i & k) 3500X graphite particles can range from few hundred to thousands of microns). The high-in plane thermal conductivity ($k \sim 2000$ Wm$^{-1}$K$^{-1}$) of graphene nanosheets forming this network, allow very efficient heat transfer over long length scales, resulting in high composite thermal conductivity. We later show through X-ray diffraction and Raman analysis that the graphitic layers forming the walls of this network, retain the chemical structure and exact interlayer spacing of graphite, and can thus be expected to have the high thermal conductivity values reported for graphite nanoplatelets ($k_{in} \sim 2000$ Wm$^{-1}$K$^{-1}$, and $k_{out} \sim 10$ Wm$^{-1}$K$^{-1}$)[12].

A unique advantage of this continuous network can be understood by noticing that even at bends/turns, the graphitic sheets are still covalently bonded, allowing for a very high thermal conductance (low interfacial thermal resistance) at such junctions. The above effect of low interfacial resistance at covalently bonded junctions in expanded graphite network, offers an advantage over the effect of percolation[13, 21, 22, 50-53] which has been shown to yield high thermal conductivity in recent years. Percolation involves enhancing heat transfer through establishing discrete particle-particle contact resulting in efficient heat conduction along a network of graphene particles. While this enhances thermal conduction by bypassing the low thermal conductivity polymer, the interfacial thermal resistance at the contact between discrete graphene particles in percolative networks can still be significant. Konatham *et al.*[54] performed molecular dynamics simulations to show that thermal contact resistance at graphene-graphene contact is $5.5 \times 10^{-8}$ K/W, which is comparable to graphene-polymer interfacial thermal resistance[8]. In expanded graphite composites, however, the covalently bonded nature of graphene sheets at bends/turns, results in much more efficient interfacial thermal transport at junctions, resulting in a superior thermal conductivity enhancement overall. Another advantage of expanded-graphite mediated *k*-enhancement over percolation effect, is that typically very high volume concentrations are required to achieve particle-particle contact for percolation. In case of solvent-cast expanded-graphite/polymer composites, however, the continuous graphitic networks are present at all volume fractions, enabling achieving high *k* even at low particle concentrations.

Polyetherimide and epoxy, the polymers chosen for this work, also enable the advantage of allowing superior thermal interaction with graphene. Both polyetherimide and epoxy have oxygen groups in their molecular structure which can enable strong thermal interaction through

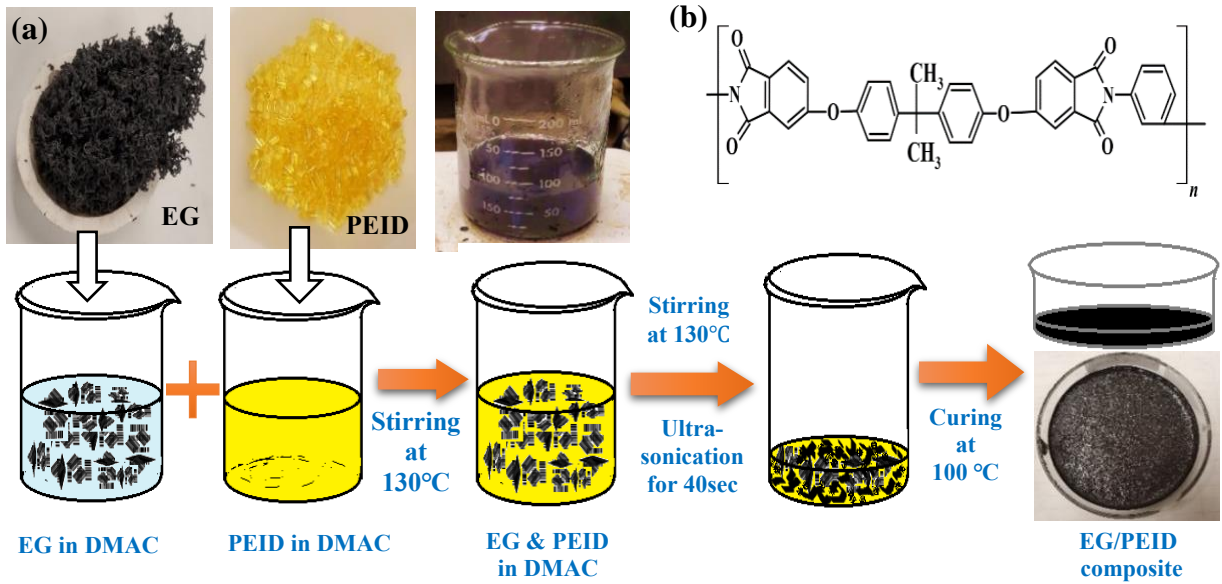

Figure 2 a) Schematic for the preparation process of EG/PEID composites, b) chemical structure of PEID

hydrogen bonding with the oxygen groups in expanded-graphene. Such oxygen groups are introduced in expanded graphite during intercalation (with an oxidizing agent) and expansion process. Evidence for presence of oxygen groups in expanded graphite is provided through X-ray photoelectron spectroscopy (XPS).

We further investigated the effect of sonication parameters on thermal conductivity enhancement and show that lower sonication time and power result in larger expanded graphite particles in the composite, which allows for heat to be conducted efficiently along longer lengths scales, resulting in higher thermal conductivities. Finally, we also compare the measured thermal conductivity results with a recently introduced effective medium model that takes graphene-graphene contact into account for thermal conductivity prediction.

## 2. Experimental Details
## 2.1. Materials

Graphite intercalated compound (GIC) or expandable graphite (EPG), with an average particle size of $\geq 180$ $\mu m$ ( +100mesh size: ~92%) and expansion ratio 290:1 (ASB-3570) were bought from Graphite store[55]. Graphene Nanoplatelets AO-4 (60 nm thickness & lateral size <7 μm ) were acquired from Graphene supermarket[56]. Epoxy resin used for the study was EPIKOTE RESIN

MGS RIMR 135 and the hardener used was EPIKURE CURING AGENT MGS RIMH 137, both purchased from Hexion[57]. Commercially available N, N-dimethylacetamide (DMAc) and polyetherimide (PEID) with melt index of 18 g/10 min (337 °C/6.6kg) and a density of 1.27 g/mL were obtained from Alfa Aesar[58] and Sigma Aldrich[59]. The organic solvents N-N, Dimethylformamide (DMF) and Acetone were purchased from University of Oklahoma chemical stock room.

## 2.2. Fabrication of the EG/Polymer composites

The fabrication procedure of the EG/PEID composite is illustrated in Fig. 2a. For EG/PEID composite preparation, required quantity of expandable graphite (0.125gm expandable graphite for 5wt% filler content EG/PEID composite of weight 2.5gm) was firstly placed into a furnace at 900°C for approximately 30-60 sec in a crucible. EG so obtained was then dispersed into 20 mL DMAc. Separately PEID pellets were dissolved into 50 mL DMAc at 130 °C for 1 h. The DMAc solution with EG was mixed with polymer solution and blended for 3 h at 130 °C followed by a short time (~ 40 sec) probe sonication at 20% amplitude. The EG blended with polymer solution of ~25-30ml was cast into a petri dish. Lastly, the petri dish was kept at 100 °C for 24-48 h to produce the composite film. Likewise, 2.5 wt%, 7.5 wt%, and 10 wt% EG/PEID composite films were prepared using this same procedure. For comparison, the graphene-nanoplatelet (GnP)/PEID composite films were also prepared using the same solution casting technique for graphene with 60 nm thickness of respectively.

To prepare epoxy/expanded graphite composites, resin was added to 90 mL N-N, dimethylformamide (DMF) solution and stirred while heating at 150°C to obtain a homogeneous mixture. Expanded graphite was added to this solution and stirred for one hour. The solution was then tip sonicated for 40 seconds followed by stirring at 150°C until the solvent completely evaporated. After solvent evaporated, a thick mixture of EG/epoxy was obtained. This mixture was then spread over a PTFE sheet and kept in a vacuum oven at 140°C for 15 h to remove any residual solvent present in the mixture. On cooling the mixture to room temperature, hardener was added to it and mixed to obtain a homogenous viscous paste. This paste was then transferred to aluminum molds and cured at 90°C for 20 h.

## 3. Characterization

**Thermal Conductivity ($k$):** $k$ of EG/polymer composites was measured by the laser flash technique. A Netzsch LFA 467 Hyperflash (Netzsch, Germany) laser was used to measure the through-thickness thermal diffusivity of the samples. 8-12 samples of 12.5mm diameter and 0.3-0.4 mm thickness were used to measure the thermal diffusivity ($\alpha$) at room temperature (23 °C). The samples were coated graphite spray before the measurement to efficiently absorb heat from a flash lamp, and an average of 6-8 measurements was reported. This laser flash technique induces heat by a laser pulse on one surface of the sample and the temperature rise is captured on the other surface of sample as a function of time. $\alpha$ is determined by LFA using the following equation: $\alpha = (0.1388 \, d^2)/t_{1/2}$, where, $\alpha$ is the thermal diffusivity (mm$^2$/s), $t_{1/2}$ is the time to obtain half of the maximum temperature on the rear surface, and d denotes the sample thickness (mm). The thermal conductivity was calculated using $k = \alpha \times \rho \times C_p$, where k, $\rho$, and Cp represent the thermal conductivity, density, and specific heat constant of the sample, respectively. In this work, density and specific heat of the composite samples were calculated using gas pycnometer (AccuPyc II 1340, Micromeritics Instrument Corporation, USA) and differential scanning calorimetry (DSC) (DSC 204F1 Phoenix, Netzsch, USA).

**Scanning Electron Microscopy (SEM):** Morphological characterization of EG filler and EG/polymer composites was carried out by high resolution Field Emission Environmental Scanning Microscopy (Quattro S FE-ESEM, Thermofisher Scientific, USA). This SEM was operated in secondary electron (SE) mode at an accelerating voltage of 20 kV. To prepare the samples for SEM imaging, liquid nitrogen was used to crack down the composite to image over the cross-sectional area. Montage large area mapping of EG fillers has been captured using MAPS software of FE-ESEM.

**Raman Spectroscopy (RS):** Raman spectroscopy (RS) was performed using a DXR3 SmartRaman Spectrometer (Thermofisher Scientific, USA) to collect the data over the range from 3250 to 250 cm$^{-1}$, laser wavelength $\lambda_L$ = 633 nm, spectral resolution = 0.16 cm$^{-1}$, and imaging resolution = 702 nm for the EG and GIC samples. An Olympus BX 41 microscope with 5x objective, 10 s exposure time for 15 accumulations, and 3 scans per sample were used to collect the spectra.

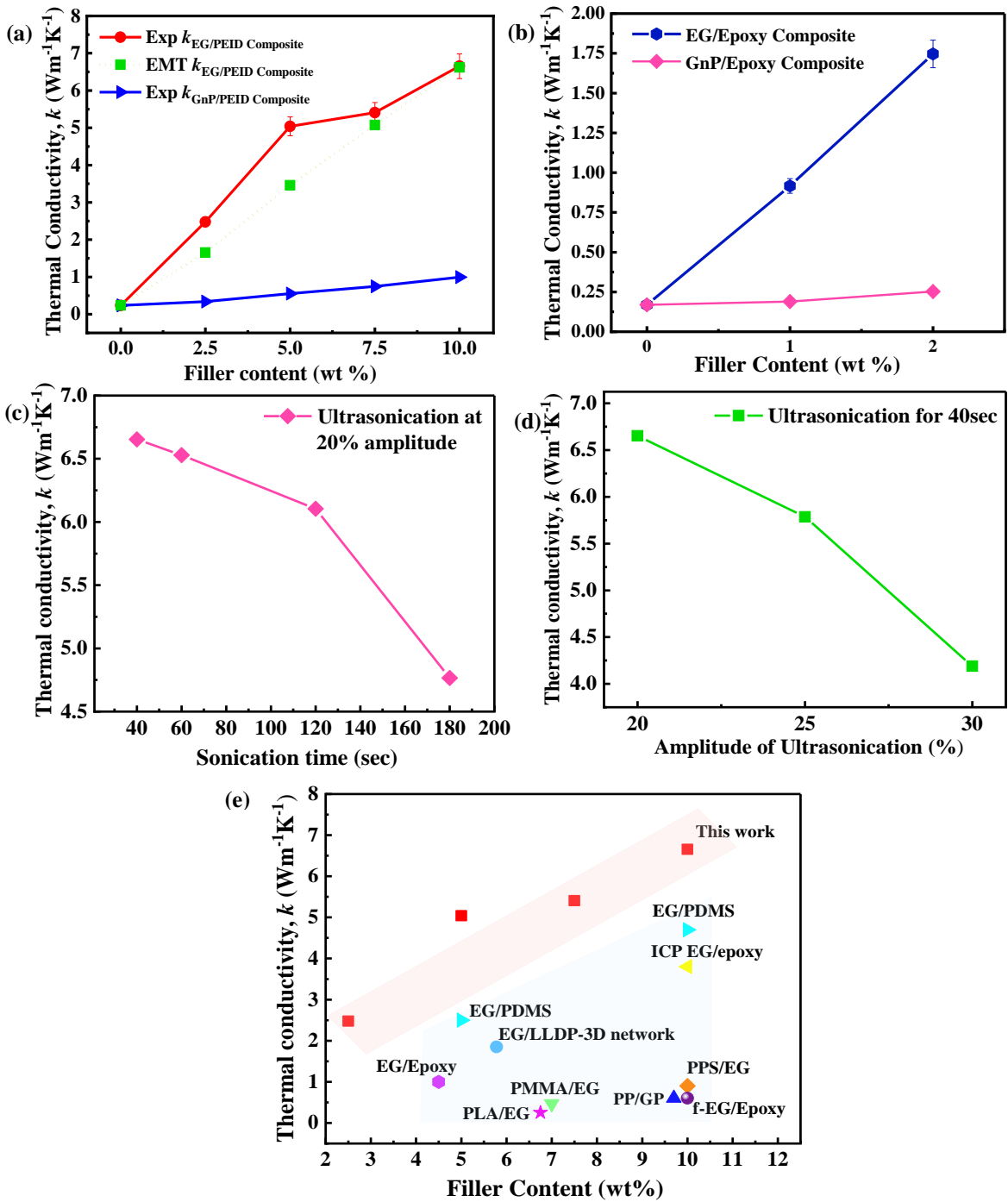

Figure 3a) Thermal conductivity value of EG/PEID (experimental & effective medium theory (EMT) )& GnP/PEID composite with different filler content (2.5,5,7.5 &10wt%), b)Thermal conductivity value of EG/epoxy composite with different filler content (0.5,1,1.5 &2wt%); c) $k$ value of EG/PEID composite with different sonication time at 20% ultrasonication power; d) $k$ value of EG/PEID composite with different ultrasonication power sonication for 40sec; e) Comparison of thermal conductivity value of polymer composites in previous works.

**X-ray Diffraction (XRD):** A PANalytical Empyrean Diffractometer (Malvern Panalytical Ltd, UK) produced the information regarding crystal structure of EG compared to GIC using Bragg-Brentano focusing geometry at room temperature. 3kW Cu Kα radiation (λ = 1.5406 Å) with a scan range of 2θ = 5 to 80° and step size of 0.013°.

**X-ray Photoelectron Spectroscopy (XPS):** X-ray Photoelectron Spectroscopy (XPS) was performed for GIC and EG sample by Thermo Scientific K-alpha XPS. Al Kα gun source was used to excite the sample and the data was collected for acquisition time of ~70 s at 400 μm spot size.

The passing energy of 200 eV was utilized to find the carbon (C), oxygen(O) & sulfur (S) peak in this analysis spectrum. The atomic percentage of C, O & S were investigated using the Avantage software. To determine the atomic percentage, this software was used to do the curve fitting in accordance with Gaussian and Lorentzian functions.

## 4. Result and Discussion
### 4.1 Thermal Conductivity Results:

The measured thermal conductivities of EG/PEID and EG/epoxy composites are shown in Figs. 3a and b. Thermal conductivity of solvent cast EG/PEID composites is measured to be around 6.6 $Wm^{-1}K^{-1}$ at 10 weight% filler composition. This value represents a remarkable 2770% enhancement over the $k$ (0.23 $Wm^{-1}K^{-1}$) of pristine PEID, providing new avenues for high thermal conductivity composites. Similarly, the measured thermal conductivity of EG/epoxy composites also shows remarkable enhancement at very low loadings of EG. At just 2 weight% EG composition, a thermal conductivity value of 1.74 $Wm^{-1}K^{-1}$ is achieved for the epoxy composite, representing an enhancement of 1025% over pristine epoxy (0.16 $Wm^{-1}K^{-1}$). Table 1 and Fig. 3e show that the measured value is significantly higher than similar graphene/polymer composites either based on a) uniformly dispersed graphene, b) other expanded graphite-based methods and c) graphene based percolative networks.

We first compare the measured values against previous results reported for percolation effect based on establishing graphene-graphene network. A key advantage of expanded-graphite is that large enhancement in $k$ value is achieved even at low graphene loading, as opposed to the case of percolation, where typically much higher particle concentrations are required to achieve particle-particle contact. Kargar *et al*. achieved around 6 $Wm^{-1}K^{-1}$ in graphene/epoxy composites

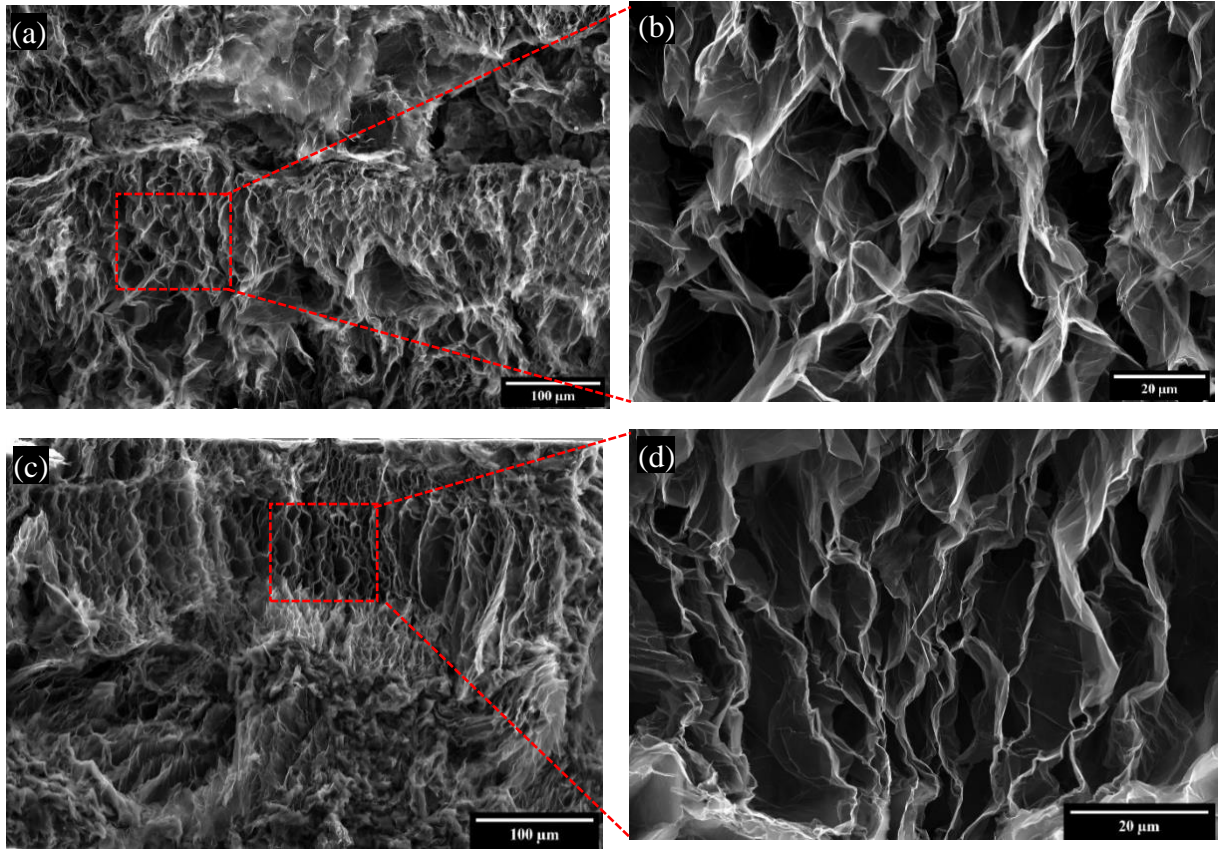

Figure 4 FE-ESEM images of EG/PEID composites with a,b) 7.5wt%, c,d) 10wt% fabricated at 20% amplitude for 40sec ultrasonication; (a,c) 350×magnification, (b,d,) 3500×magnification

at a high graphene loading of 35 vol% through percolation effect[21]. Percolation was also found to yield a thermal conductivity of ~ 5.5 Wm$^{-1}$K$^{-1}$ in Boron-Nitride/epoxy composites[21] at a volume loading of 45 vol%. High $k$ of 6.6 Wm$^{-1}$K$^{-1}$ is achieved in this work through use of just 10 weight% EG content. Even in percolative $k$ enhancement, the thermal interfacial resistance at particle-particle contact can be relatively high. Konatham *et al*. reported a thermal boundary resistance of $5.5 \times 10^{-8}$ K/W, almost as high as at the graphene-polymer contact[54]. Expanded graphite, achieved through solvent casting approach, leads to continuous graphite networks, overcoming the issue of low particle-particle interfacial thermal conductance in percolative environments.

High $k$ achieved in this work is also due to the use of solvent casting, which offers the advantage of preserving the porous structure of expanded graphite within the composite (seen in Figs. 1f, i and k ). This is due to only moderate forces being exerted on expanded graphite during solvent casting approach. Figs 1f and i compare the porous structure of expanded graphite, before

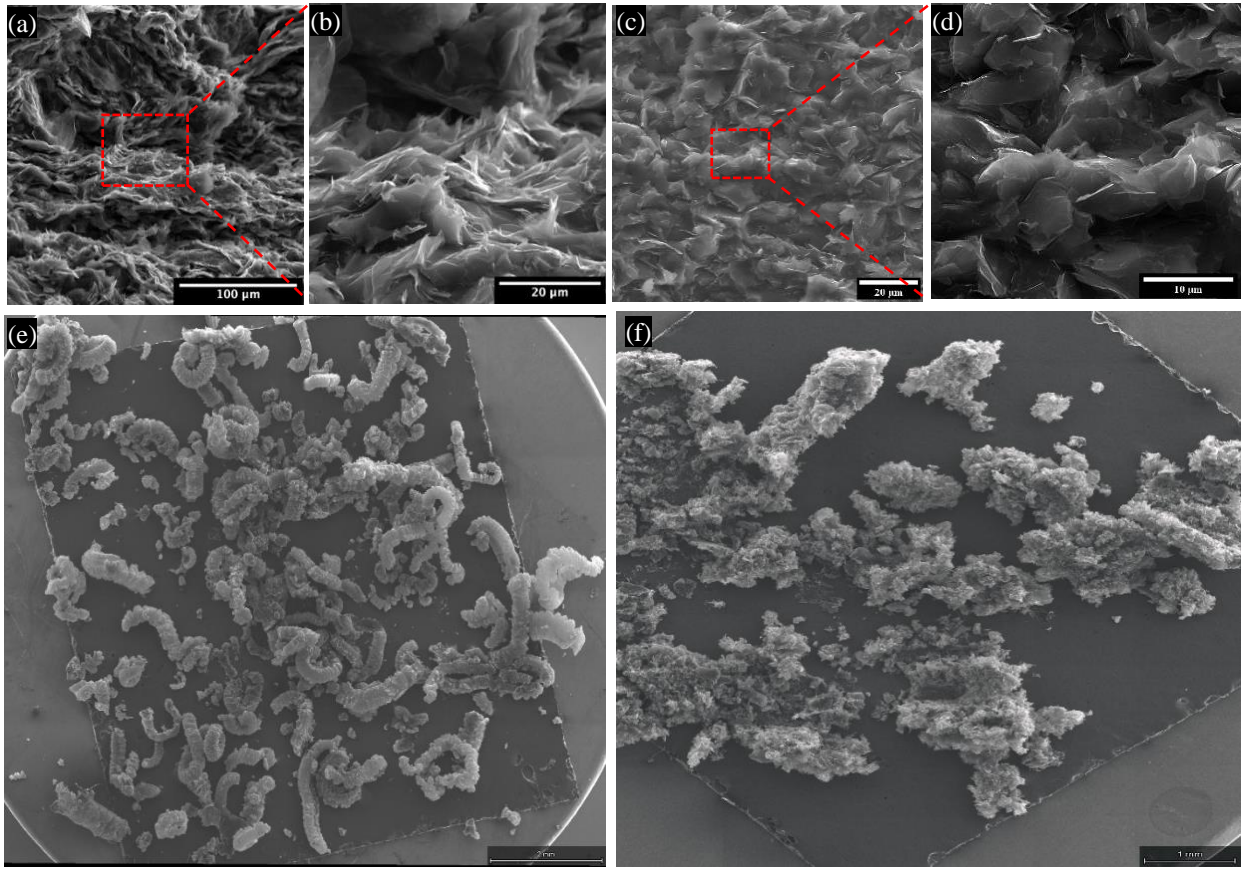

Figure 5 FE-ESEM images of (a,b)10 wt% EG/PEID composite with ultrasonication for 3minutes at 20% amplitude, (c,d) 10wt% GnP/PEID composite with ultrasonication for 40sec at 20% amplitude; FE-ESEM images of EG at 20% sonication for e) 40 sec and f) 3 min using montage large area mapping

and after it is embedded into the composite, and show that the porous structure of EG is largely retained within the composite. This is in contrast to micro-compounding (melt-blending) where large shear forces during the compounding process, can exfoliate expanded graphite, converting it into a nanoplatelet like morphology. Mokhtari *et al.*[60] discussed such exfoliation of expanded graphite through microcompounding.

We also compare measured $k$ values of EG/PEID composite with those of GnP/PEID composites in Figs. 3a and b. The measured thermal conductivity of EG/PEID composites is dramatically higher relative to that achieved using graphite nanoplatelets. At 10 wt% composition, the $k$ of EG/PEID composite (6.6 Wm$^{-1}$K$^{-1}$) is higher by 572% relative to GnP/PEID composite (1.0 Wm$^{-1}$K$^{-1}$). Similarly, at 2 wt% composition, $k$ of EG/epoxy composite (1.74 Wm$^{-1}$K$^{-1}$)) is higher by ~600% relative to GnP/epoxy composite (0.25 Wm$^{-1}$K$^{-1}$). At the low graphene content used in this work, GnPs are well separated by polymer; high interfacial thermal resistance between GnPs and polymer then restricts the enhancement achievable through use of graphene nanoplatelets. These results highlight the large advantage of expanded graphite over graphene nanoplatelets in enhancing composite thermal conductivity.

We further investigated the effect of sonication parameters used during composite preparation on $k$ enhancement. Figs. 3c and d show that lower sonication time and amplitude leads to higher thermal conductivity of EG/PEID composite. To understand the effect of sonication parameters on structural integrity of EG, we performed FE-SEM analysis. As the porous interpenetrating network has beneficial impact on thermal conductivity enhancement (shown in fig 4a-d), high resolution images are obtained to visualize the effect of sonication time on this porous network structure (Fig 5a, b). The image of 10 wt% EG/PEID composite (Fig. 5a and b) prepared with 3min sonication time at 20% amplitude shows an absence of porous structure within the composite suggesting that longer sonication time causes the porous structure to be damaged. FE-SEM image of 10wt% GnP/PEID composite are also presented here in Figs. 5c & d.

While above images show the expanded graphite embedded in polymer, we also show expanded graphite before it is embedded into the polymer in Figs. 5e & f. These images show that fragile, porous graphite is broken into nanosheets after 3 minutes sonication time whereas 40 sec sonication time has negligible effect on EG filler structure. Short time sonication thus offers the advantage of preserving the structure of expanded graphite, while also allowing uniform dispersion into the polymer matrix.

In the next section, we compare measured $k$ values of EG/PEID composite with theoretical predictions, based on a recently introduced effective medium model by Su *et al.*[61]. Comparison with theoretical predictions highlight the advantage of expanded graphite and provides evidence for the outlined mechanism of heat conduction along continuous graphitic paths.

**Table 1 Comparison of $k_\perp$ for different polymer-graphene and EG-polymer composite**

| Filler | Matrix | Fraction | $k_\perp$ (Wm$^{-1}$K$^{-1}$) | Enhancement (%) | Preparation method | ref |
|---|---|---|---|---|---|---|
| GnPs/MWCNT | PS | (3.5/1.5) vol% | 1.02 | 437 | Melt mixing + Synergistic effect | 62 |
| Graphite | PP | 40wt% | 5.4 | 2060 | Melt mixing, compression molding | 63 |
| Multilayer GnP | Epoxy | 10vol% | 5.1 | 2300 | Solvent casting, higher sheer mixing | 8 |
| fGO | Epoxy | 5wt% | 0.21 | 34 | Solution casting | 64 |
| EG | LLDP | 5.78wt% | 1.85 | 461 | Melt mixing, 3D network formation | 49 |
| EG | PMMA | 7 wt% | 0.47 | 276 | water-assisted melt extrusion | 65 |
| EG | LDPE | 10 wt% | 0.5 | 56 | Melt mixing | 66 |
| EG | PDMS | 10 wt% | 4.7 | 2511 | Solvent casting, hot press | 46 |
| EG | PEG | 10 wt% | 1.324 | 344 | Melt mixing | 67 |
| EG | Paraffin | 25 wt% | 3.16 | 1695 | Melt mixing | 47 |
| EG | PEID | 30 wt% | 1.6 | 700 | Solvent mixing, Melt mixing followed by injection molding | 68 |
| EG | PEID | 10 wt% | 6.65 | 2770 | Solution casting | This work |

PS-polystyrene; PP-polypropylene; PMMA- Poly(methyl methacrylate); LDPE- Low density polyethylene; PDMS-polydimethylsiloxane; PEG- Polyethylene glycol

## 4.2 Effective medium Model for TC

Effective medium model introduced by Su *et al.*[61] includes the effect of both graphene-polymer and graphene-graphene thermal contact resistance. Effective composite thermal conductivity $k_e$ through this model is computed by solving the following equation,

$$c_0 \frac{k_0 - k_e}{k_e + (k_0 - k_e)/3} + \frac{c_1}{3} \left[ \frac{2(k_{11} - k_e)}{k_e + S_{11}(k_0 - k_e)} + \frac{(k_{33} - k_e)}{k_e + S_{33}(k_3 - k_e)} \right] = 0$$

where $c_0$ and $c_1$ are the concentrations of the embedding matrix and filler material, respectively, $k_{11}$ and $k_{33}$ are the effective in-plane and through-plane thermal conductivities of graphitic nanosheets. The effective thermal conductivities take into account the effect of interfacial thermal resistance. $S_{11}$ and $S_{33}$ are the shape parameters related to the aspect ratio of graphitic nanosheets, given by following equations.

$$S_{11}= S_{22} = \frac{\alpha}{2(1-\alpha^2)^{3/2}}\left[\arccos\alpha - \alpha(1-\alpha^2)^{1/2}\right], \quad \alpha<1$$

$$S_{33} = 1 - S_{11}$$

In above equations, $\alpha$ is the aspect ratio (thickness/lateral dimension) of the graphitic nanosheets and $k_0$ is thermal conductivity of an interlayer surrounding graphene sheets. The role of this interlayer is to include the effect of interfacial thermal resistance at graphene-polymer and graphene-graphene contacts.

The effective thermal conductivities $k_{11}$ and $k_{33}$ are computed using the in-plane and through-plane thermal conductivities of graphene ($k_1$ and $k_3$ respectively) and the interlayer properties through the following equations,

$$k_{11} = k_0\left[1 + \frac{(1-c_{int})(k_1-k_0)}{c_{int}S_{11}(k_1-k_0)+k_0}\right] \qquad (15)$$

$$k_{33} = k_0\left[1 + \frac{(1-c_{int})(k_3-k_0)}{c_{int}S_{11}(k_3-k_0)+k_0}\right] \qquad (16)$$

In above equations, $c_{int}$ is the concentration of the interlayer. The values of different parameters used in this effective medium model are described below.

Table 2 below shows the values of different parameters used in above calculations.

| Material Parameters | Values |
|---|---|
| Average graphene lateral length, $l$, | 10 μm |
| Average graphene thickness | 10 nm |
| Aspect ratio of the graphene filler | 0.001 |
| Thermal conductivity of polymer phase | 0.23 Wm$^{-1}$K$^{-1}$ |
| Thermal conductivity of graphene filler, $k_1$ and $k_3$ (Wm$^{-1}$K$^{-1}$) | 2000 and 10 Wm$^{-1}$K$^{-1}$ |
| Thermal conductivity of interlayer with Kapitza resistance | 0.04 Wm$^{-1}$K$^{-1}$ |
| Thermal conductivity of the interlayer with a firmly developed graphene-graphene contact state, | 0.17 Wm$^{-1}$K$^{-1}$ |

In above table, graphene thickness and length are the thickness and length of the graphitic sheets forming the walls of the interpenetrating network. Average values of these parameters were obtained from microscopy to be 10 μm for lateral length and 10 nm for thickness. Thermal conductivity of the polymer phase is taken to be 0.23 Wm$^{-1}$K$^{-1}$ from our measurements (in good agreement with literature). The in-plane ($k_1$) and through-plane thermal conductivities ($k_3$) of graphene were taken to be 2000 Wm$^{-1}$K$^{-1}$ and 10 Wm$^{-1}$K$^{-1}$ respectively[12, 69].

The interfacial resistance between graphene and polymer is modeled as an interlayer in above theory. The thickness of this interlayer was nominally taken to be 2 nm and its thermal conductivity was assumed to be 0.04 Wm$^{-1}$K$^{-1}$, resulting in an interfacial resistance between graphene and polymer of $5 \times 10^{-8}$ m$^2$K/W[22], a well-accepted value for interfacial thermal resistance between graphene and polymer.

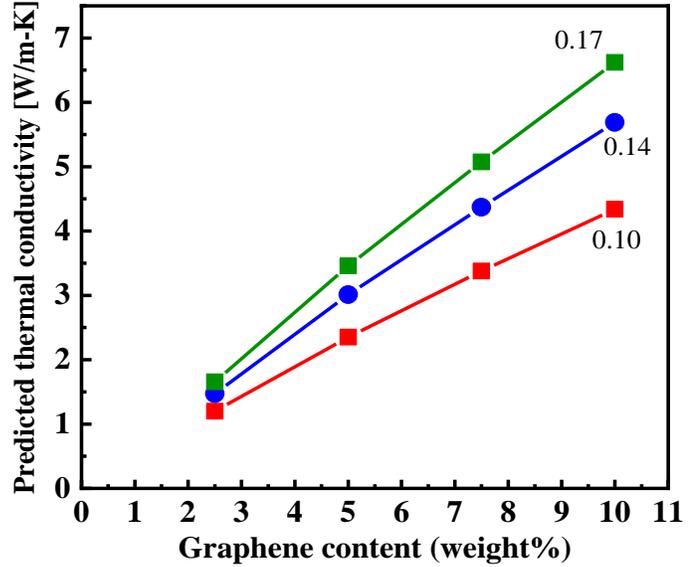

Figure 6 Predicted thermal conductivity value based on the effective medium theroy (EMT) with different graphene-graphene interlayer thermal conductivity.

A key parameter in above model is the interfacial resistance at graphene-graphene contact. This value was determined by fitting the measured values to the effective medium model. A good agreement between measured thermal conductivities and predicted values was obtained by using a graphene-graphene interlayer thermal conductivity of 0.17 Wm$^{-1}$K$^{-1}$ (see Fig. 3a and Fig. 6), which is more than 4 times higher than the interlayer thermal conductivity of polymer-graphene (0.04 Wm$^{-1}$K$^{-1}$). This interlayer thermal conductivity of 0.17 Wm$^{-1}$K$^{-1}$ at graphene-graphene contact corresponds to a graphene-graphene interfacial thermal resistance of $1.17 \times 10^{-8}$ m$^2$K/W. This value is significantly smaller than the interfacial resistance predicted for regular graphene-graphene contact[54] ($5 \times 10^{-8}$ m$^2$K/W) providing evidence that graphene sheets are in superior contact (covalently bonded) at bends/turns in expanded graphite network, compared to the contact between discrete graphene particles. The analysis points to the advantage of the continuous graphitic network achieved through the use of expanded graphite via solvent casting technique in enhancing thermal conductivity of polymer-composites. We further show the effect of lower graphene-graphene contact conductance by decreasing the graphene-graphene contact thermal conductivity in the model. It is seen that as the graphene-graphene contact thermal conductivity is decreased from 0.17 to 0.10, the effective composite thermal conductivity is decreased from 6.6 Wm$^{-1}$K$^{-1}$ to 4.4 Wm$^{-1}$K$^{-1}$ (by almost 33.3%), indicating the importance of superior graphene-graphene contact conductance for achieving overall high composite $k$ values. We next discuss the characterization of EG filler and EG/PEID composites.

### 4.3 Morphologies of EG filler and EG/PEID composite:

Analysis through FE-ESEM reveals the structural integrity of EG structure before and after the preparation of polymer composite. SEM (Scanning electron microscopy) images enable understanding of morphological differences between graphene-nanoplatelets (GnPs), Graphene-Intercalation compound (GIC) and thermally expanded graphite (EG) (Fig 1a). Thermal expansion of GIC with average diameter of ~180 μm and thickness of 1-150 μm turns it into worm like structure. Uneven expansion resulting from expansion of intercalation compound at 900 °C, leads to the separation of expandable graphite into multiple layers resulting in a porous network with average edge size of 10-20μm. These pores allow the formation of interpenetrating graphene-polymer network where the pores are wetted with PEID polymer. While Fig. 1f shows expanded graphite before it is used to prepare composite, FE-SEM also is used to study the structure of EG even within the polymer composite. EG structures in EG/PEID composites with 7.5 & 10wt% filler are presented in Figs. 4a-d. These figures show that solvent casting clearly preserves the porous network structure of expanded graphite, enabling creation of the highly thermally conductive interpenetrating graphene-polymer network within the composite.

### 4.4 Analysis of Crystal structure by XRD and Raman Spectroscopy

X-ray diffraction (XRD) analysis was performed to determine the crystal structure and interlayer spacing of GIC and EG. Fig. 7a shows a strong diffraction peak at 2θ= 26.133° (002) for GIC, slightly shifted from the case of natural graphite 2θ= 26.5°[70-73]. A weaker peak (004) is observed at 2θ=54.37° [74] for GIC. The small shift in peak for GIC is attributed to the presence of intercalated compounds. On the contrary, a reduced sharp peak is visible at 2θ= 26.35° (002)[75] for EG (the inset of Fig. 7a), closer to the (0 0 2) graphitic carbon structure. A clear diminution is observed in the intensity at (0 0 2) peak which is caused due to disorder in graphitic morphology [76] after expansion process. Still, a mostly aligned peak position in EG indicates the existence of intact chemical structure of graphite and interlayer order[77, 78]. This interconnected and stacked structure of EG enables better thermal transport throughout the polymer composite[26].

Nondestructive Raman analysis was also performed to further analyze the crystal structure before and after the thermal expansion. Raman spectroscopy of Fig. 7b exhibits two inherent peaks of G-band and D-band at ~1580 cm$^{-1}$ and ~1350 cm$^{-1}$ for graphitic material[79]. The G band signifies the stretching of defect-free sp$^2$ carbon of hexagonal ring due to in-plane tangential stretching of the carbon-carbon bonds[80] and the D band represents the vibrational mode, caused by the amorphous disordered structure of sp$^3$ hybridized carbon[81]. 2D band also can be seen at around 2700 cm$^{-1}$ and represents a second-order two photon process[82]. Raman spectra of GIC clearly shows those characteristic peaks (G, D & 2D bands). In contrast, G and 2D bands are present but presence of D band is negligible in EG Raman spectra. $I_D/I_G$ ratio is typically used to characterize the defective state of graphene. The absence of a D band in EG suggests the presence of a highly ordered defect-free graphite structure in EG. This high degree of ordered structure of graphite even after thermal expansion has a strong beneficial impact on thermal conductivity enhancement of polymer graphene composite as it preserves the intrinsic high thermal conductivity of graphene itself.

## 4.5 XPS analysis of GIC and EG

**Table 3 Atomic composition by XPS analysis of EG and GIC**

|  | Atomic Composition by XPS (at%) | | |
| --- | --- | --- | --- |
|  | C (285.08 eV) | O (532.08 eV) | S (169.11eV) |
| GIC | 85.14 | 13.16 | 1.7 |
| EG | 95.76 | 4.24 | - |

XPS analysis was further performed to investigate the concentration of carbon (C), oxygen (O) and sulfur (S) elements before and after the thermal expansion of GIC as presented in table 3. Fig. 7c shows that two peaks of C1s and O1s at ~285eV and ~532eV are present for GIC and EG filler but S2p peak (~169eV) is only visible in GIC spectra because of the included intercalated compounds. Atomic percentage of carbon increases from 85.14% to 95.76% and oxygen reduces from 13.16% to 4.24% after expansion. This is attributed to the fact that thermal expansion at 900 °C releases the oxygen contents. The presence of small amounts of oxygen groups in EG allows favorable interactions with oxygen groups in PEID through hydrogen bonding, leading to efficient interfacial thermal transport. This further enhances thermal conductivity of EG/PEID composite.

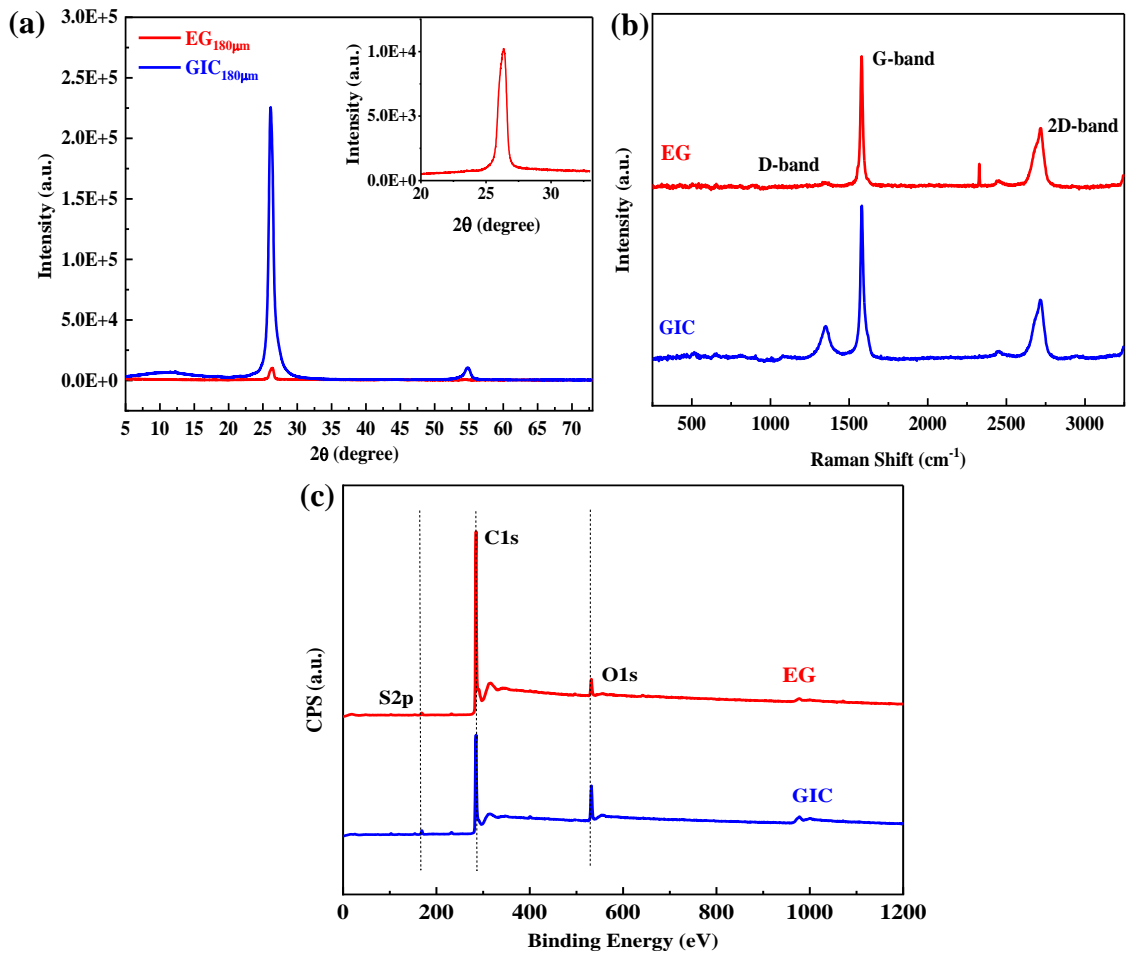

Figure 7 XRD spectra (a), RAMAN spectra (b), and XPS spectra (c) of GIC and EG filler

## 5. CONCLUSION

In summary, we demonstrate that expanded graphite (EG) can lead to a large enhancement in thermal conductivity of EG/PEI composites prepared through solvent casting. At 10 wt% EG composition, a high thermal conductivity of 6.6 $Wm^{-1}K^{-1}$ is measured, representing an enhancement of 2770% over pristine polyetherimide. This large enhancement in thermal conductivity is found to be a due to a network of continuous graphene sheets over long length scales achieved through solvent casting technique which preserves the interconnected porous structure of expanded graphite within the composite. Even at bends/turns in graphene sheets in such a network, the sheets are covalently bonded which minimizes the interfacial thermal resistance at junctions, enhancing heat transfer. Overall, the resulting structure allows highly efficient heat conduction over long length scales along the continuous graphitic sheets. This results in the observed high thermal conductivity of the composite through use of EG. Thermal conductivity of EG/PEID composite is also found to dramatically exceed that of graphene-nanoplatelet (GnP)/PEID composites by 572% for 10 wt% filler composition. At low filler loading, GnPs are well separated by polymer, and the resulting high graphene-polymer interfacial thermal resistance, results in low effective GnP/PEID thermal conductivity. Presented results highlight the advantage of expanded graphite in enhancing thermal conductivity of polymer composites, and can lead to novel avenues for achieving efficient thermal management in a wide array of technologies.

## Acknowledgment

FT, SD, JG and AN acknowledge support from National Science Foundation CAREER award under Award No. #1847129. We also thanks Mohammed Ibrahim, PhD, from Thermo Fisher Scientific for collecting the Raman spectra.

## Conflicts of interest

There are no conflicts to declare.